\documentclass[11pt]{article}
\usepackage{jheppub}

\newcommand\fft[2]{{\frac{#1}{#2}}}
\newcommand\ft[2]{{\textstyle\frac{#1}{#2}}}
\newcommand\nn{\nonumber}

\begin{document}

\preprint{MCTP-13-39}

\title{\boldmath The shortened KK spectrum of IIB supergravity on $Y^{p,q}$}

\author[a]{Arash Arabi Ardehali,}
\author[a]{James T. Liu,}
\author[b]{and Phillip Szepietowski}

\affiliation[a]{Michigan Center for Theoretical Physics, Randall Laboratory of Physics,\\
The University of Michigan, Ann Arbor, MI 48109--1040, USA}
\affiliation[b]{Department of Physics, University of Virginia,\\
Box 400714, Charlottesville, VA 22904, USA}

\emailAdd{ardehali@umich.edu}
\emailAdd{jimliu@umich.edu}
\emailAdd{pgs8b@virginia.edu}

\abstract{We examine the shortened KK spectrum of IIB supergravity
compactified on $Y^{p,q}$ and conjecture that the spectrum we have
obtained is complete. The (untwisted) shortened spectrum on
$S^{5}/\mathbb{Z}_{2p}$ and on $T^{1,1}/\mathbb{Z}_{p}$ are obtained as
special cases when $p=q$ and $q=0$, respectively. Knowledge of the
shortened spectrum allows us to compute the superconformal index of
these theories and to find agreement with earlier calculations from the
dual field theories. We also employ the shortened spectrum to
perform a $1/N^{2}$ test of AdS/CFT by holographically reproducing
the difference of the central charges, $c-a=p/8$, of the dual CFTs.}

\maketitle
\flushbottom

\section{Introduction}

About a decade ago a new avenue was opened in the exploration of
AdS/CFT with reduced supersymmetry by the discovery of an infinite
family of Sasaki-Einstein five-manifolds $Y^{p,q}$
\cite{Gauntlett:2004a,Gauntlett:2004b}
and the construction of their dual four-dimensional quiver gauge
theories \cite{Martelli:2004,Benvenuti:2004}. Various checks of the
duality had been performed successfully, the most notable of which
being perhaps the matching of the large-$N$ conformal anomalies and
the spectrum of baryonic states \cite{Benvenuti:2004,Canoura:2005}.

However, although many other properties
of the family were known from the field theory side, such as the spectrum of the mesonic states
\cite{Benvenuti:2005} and the superconformal index
\cite{Gadde:2011}, the gravitational computations were obstructed by
the difficulty in obtaining the KK spectrum of IIB supergravity on the
$Y^{p,q}$ manifolds. In particular, the scalar Laplacian on $Y^{p,q}$ leads to a
Heun equation \cite{Berenstein:2005xa,Kihara:2005nt,Oota:2005mr} whose exact
spectrum is not known.

Despite the difficulty in finding the full spectrum of the Heun
equation in question, some harmonics were found and identified with
their dual mesonic states in \cite{Kihara:2005nt}. Here we extend
the available results by finding the shortened KK
spectrum of IIB supergravity compactified on $Y^{p,q}$. We conjecture
that this spectrum is complete and perform two checks involving AdS/CFT.
This shortened spectrum is the main result of this work and can be found
in Tables~\ref{tbl:KKshort} and \ref{tbl:scalarshort}.

The shortened KK spectrum enables us to see the spectrum of all the
protected single-trace operators from the gravity side. It also
allows us to compute the superconformal index from supergravity and
find matching with the earlier field theoretical computation of
\cite{Gadde:2011}.

Another use of the shortened spectrum is in holographically
reproducing the difference of the gauge theory central charges $c-a$
\cite{Ardehali:2013}. This is a rather non-trivial test of
AdS/CFT beyond large-$N$. Our success in reproducing the field
theory value, $c-a=p/8$, from the shortened spectrum on the gravity
side helps to address some of the issues raised in
\cite{Ardehali:2013}.

This paper is organized as follows. Section~\ref{sec:SE5spect}
reviews the KK spectrum of IIB supergravity compactified on a
generic Sasaki-Einstein 5-manifold and the possible multiplet
shortening patterns. In section~\ref{sec:YpqSpect} we present the
shortened KK spectrum of IIB supergravity on $Y^{p,q}$.
Section~\ref{sec:Ring} illustrates how the mesonic chiral ring of
the dual field theories is mapped to the supergravity states. In
section~\ref{sec:SpecialCases} we demonstrate that the untwisted
shortened spectrum on $S^5/\mathbb{Z}_{2p}$ and
$T^{1,1}/\mathbb{Z}_{p}$ may be obtained from the shortened spectrum
presented here upon setting $p=q$ or $q=0$, respectively. The
superconformal index is computed in section~\ref{sec:Index}, and the
holographic $c-a$ is computed in section~\ref{sec:c-a}. The closing
section includes comments on how the results of the present paper
shed light on the issues concerning the $1/N^2$ corrections to the
holographic Weyl anomaly raised in \cite{Ardehali:2013}.

\section{The KK spectrum of IIB supergravity on SE$_5$}
\label{sec:SE5spect}

A generic compactification of IIB supergravity on AdS$_5\times\mathrm{SE}_5$
yields $\mathcal N=2$ gauged supergravity coupled to a KK tower that can be
arranged into $\mathcal N=2$ representations of the $\mathrm{SU}(2,2|1)$
supergroup.

The compactification on $S^5$ preserves $\mathcal N=8$
supersymmetry, and the KK spectrum was obtained in
\cite{Gunaydin:1984fk,Kim:1985ez}.  The result is particularly
simple when given in terms of shortened representations of
$\mathrm{SU}(2,2|4)$; at level $p$ ($p\ge2$), the states transform
under the representation $\mathcal D(p,0,0;0,p,0)$, where we have
used the notation $\mathcal D(E_0,s_1,s_2;l_1,l_2,l_3)$ where
$(l_1,l_2,l_3)$ are the Dynkin labels of the $\mathrm{SU}(4)_R$
representation.

Subsequently the KK spectroscopy for $T^{1,1}$ was investigated in
\cite{Ceresole:1999zs,Ceresole:1999ht}.  The resulting spectrum was given in
terms of nine generic KK multiplets --- Graviton, Gravitinos I through IV, and Vectors I through IV ---
along with a Betti vector and Betti hypermultiplet.  It was then shown in \cite{Eager:2012hx}
that this decomposition in terms of nine generic multiplets persists for general $\mathcal N=2$
compactifications.  The full spectrum consists of these generic multiplets along with
the possible addition of special KK multiplets and Betti multiplets.

In fact, the analysis of \cite{Ceresole:1999zs,Ceresole:1999ht,Eager:2012hx} demonstrates
that the generic KK tower can be obtained solely from knowledge of the eigenvalues of
the scalar Laplacian on SE$_5$.  Essentially, the vector and tensor harmonics needed in
the decomposition of IIB fields on SE$_5$ may be related to a combination of scalar
harmonics and invariant tensors related to the structure of the manifold.  Hence information
from the scalar harmonics is sufficient.

It is convenient to define the eigenvalues of the scalar Laplacian on SE$_5$ according to
\begin{equation}
\square Y = -e_0(e_0+4) Y,
\label{eq:laplacian}
\end{equation}
where we take $e_0\ge0$.
Note that the eigenvalues $e_0$ will depend on the $R$-charge as well as other
global quantum numbers on SE$_5$.  Moreover, it was shown in \cite{Eager:2012hx}
that $e_0$ satisfies the bound
\begin{equation}
e_0\ge\ft32r.
\end{equation}
As an example, for $T^{1,1}$, we have
\begin{equation}
e_0(e_0+4)=6[j(j+1)+\ell(\ell+1)-r^2/8],
\end{equation}
where $(j,l,r)$ labels the representation under the isometry group
$\mathrm{SU}(2)_j\times\mathrm{SU}(2)_\ell\times\mathrm U(1)_r$ of $T^{1,1}$,
and the $R$-charge satisfies the bound
\begin{equation}
|r|\le2\,\mbox{min}(j,\ell).
\end{equation}
The $e_0\ge\fft32r$ bound is saturated when $j=\ell=|r|/2$.

In general, the isometry group may be different.  However, the conserved
$\mathrm U(1)_r$ will always be present, as demanded by $\mathcal N=2$
supersymmetry.  Thus the KK spectrum can be arranged into representations of
$\mathrm{SU}(2,2|1)$ based on the $e_0$ and $r$ eigenvalues of the scalar Laplacian.
The generic KK spectrum is given in Table~\ref{tbl:KKspectrum}.
In addition to the generic spectrum, there may be KK towers of special multiplets
as well as a finite number of Betti multiplets.  The former correspond
to two-forms $\omega\in H_{\bar\partial_B}^{1,1}(\mathrm{SE}_5)$, and the latter
to $\omega\in H^2(\mathrm{SE}_5)$ \cite{Eager:2012hx}.

\begin{table}[tp]
\centering
\begin{tabular}{|l|l|l|}
\hline
Supermultiplet&Representation&$e_0$ condition\\
\hline
Graviton&$\mathcal D(e_0+3,\ft12,\ft12;r)$&$e_0\ge0$\\
Gravitino I and III&$\mathcal D(e_0+\ft32,\ft12,0;r+1)+\mathcal
D(e_0+\ft32,0,\ft12;r-1)$&
$e_0>0$\\
Gravitino II and IV&$\mathcal D(e_0+\ft92,\ft12,0;r-1)+\mathcal
D(e_0+\ft92,0,\ft12;r+1)$&
$e_0\ge0$\\
Vector I&$\mathcal D(e_0,0,0;r)$&$e_0>0$\\
Vector II&$\mathcal D(e_0+6,0,0;r)$&$e_0\ge0$\\
Vector III and IV&$\mathcal D(e_0+3,0,0;r-2)+\mathcal D(e_0+3,0,0;r+2)$&$e_0\ge0$\\
\hline
\end{tabular}
\caption{\label{tbl:KKspectrum} The generic $\mathcal N=2$ spectrum of
IIB supergravity on SE$_5$.  The spectrum is given in terms of the eigenvalue $e_0$
of the scalar Laplacian and the $R$-charge $r$.}
\end{table}

The scalar Laplacian always admits a constant mode on SE$_5$, with corresponding
eigenvalues $e_0=0$ and $r=0$.  Truncating to $e_0=0$ gives the zero mode spectrum
shown in Table~\ref{tbl:KKconsistent}.  These are the modes that may be retained in the
consistent truncation on any squashed Sasaki-Einstein manifold
\cite{Cassani:2010uw,Liu:2010sa,Gauntlett:2010vu,Skenderis:2010vz}.

\begin{table}[tp]
\centering
\begin{tabular}{|l|l|l|}
\hline
Supermultiplet&Representation&Name given in \cite{Liu:2010sa}\\
\hline
Graviton&$\mathcal D(3,\ft12,\ft12;0)$&supergraviton\\
Gravitino II and IV&$\mathcal D(\ft92,\ft12,0;-1)+\mathcal D(\ft92,0,\ft12,1)$
&LH+RH massive gravitino\\
Vector II&$\mathcal D(6,0,0;0)$&massive vector\\
Vector III and IV&$\mathcal D(3,0,0;-2)+\mathcal D(3,0,0;2)$&LH+RH chiral\\
\hline
\end{tabular}
\caption{\label{tbl:KKconsistent} The $e_0=0$ multiplets that may be retained in
a consistent truncation.}
\end{table}

\subsection{Multiplet shortening}

In general, the spectrum in Table~\ref{tbl:KKspectrum} fill out long multiplets.  However,
the multiplets will be shortened whenever some of the unitarity bounds become saturated
\cite{Flato:1983te,Dobrev:1985qv} (see also \cite{Freedman:1999gp}).  There are three
multiplet shortening conditions
\begin{eqnarray}
\mbox{conserved:}&&\qquad E_0=2+s_1+s_2,\quad \ft32r=s_1-s_2,\nn\\
\mbox{chiral (anti-chiral):}&&\qquad E_0=\ft32r\quad(E_0=-\ft32r),\nn\\
\mbox{semi-long I (semi-long II):}&&\qquad E_0=2+2s_1-\ft32r\quad(E_0=2+2s_2+\ft32r).
\end{eqnarray}
Note that the conserved multiplet can be thought of as satisfying the semi-long I and II conditions
simultaneously.

We now impose these shortening conditions on the generic KK spectrum in
Table~\ref{tbl:KKspectrum}.  Making note of the restriction $e_0\ge\fft32|r|$, we find that
shortening occurs only under the conditions
\begin{equation}
i)\quad e_0=\ft32|r|\qquad\mbox{or}\qquad ii)\quad e_0=\ft32|r|+2.
\label{eq:shortc}
\end{equation}
These two possibilities were noted in \cite{Ceresole:1999zs,Ceresole:1999ht} in the
case of $T^{1,1}$.
The possible shortenings are given in Table~\ref{tbl:KKshort}.  Note that $e_0=\fft32r$
($e_0=-\fft32r$) corresponds to holomorphic (antiholomorphic)  functions on the
Calabi-Yau cone over SE$_5$ \cite{Eager:2012hx}.

\begin{table}[tp]
\centering
\begin{tabular}{|l|l|l|l|l|}
\hline
Multiplet&Representation&\multicolumn{2}{c|}{Shortening condition}&Shortening type\\
\hline
Graviton&$\mathcal D(e_0+3,\fft12,\fft12;r)$&$e_0=0$&$r=0$&conserved\\
&&$e_0=-\fft32r$&$r<0$&SLI\\
&&$e_0=\fft32r$&$r>0$&SLII\\
\hline
Gravitino I&$\mathcal D(e_0+\fft32,\fft12,0;r+1)$&$(e_0=1)$&$(r=-\fft23)$&(conserved)\\
&&$e_0=\fft32r$&$r>0$&chiral\\
&& $e_0=-\fft32r$&$r<-\fft23$&SLI\\
&& $e_0=\fft32r+2$&$r>-\fft23$&SLII \\
\hline
Gravitino II&$\mathcal D(e_0+\fft92,\fft12,0;r-1)$&$e_0=-\fft32r$&$r\le0$&SLI\\
\hline
Gravitino III&$\mathcal D(e_0+\fft32,0,\fft12;r-1)$ & $(e_0=1)$&$(r=\fft23)$&(conserved)\\
&& $e_0=-\fft32r$&$r<0$&anti-chiral\\
&& $e_0=\fft32r$&$r>\fft23$&SLII\\
&& $e_0=-\fft32r+2$&$r<\fft23$& SLI\\
\hline
Gravitino IV &$\mathcal D(e_0+\fft92,0,\fft12;r+1)$& $e_0=\fft32r$&$r\ge0$&SLII\\
\hline
Vector I &$\mathcal D(e_0,0,0;r)$&$e_0=2$&$r=0$&conserved\\
&&$e_0=\fft32r$&$r\ge\fft23$&chiral\\
&&$e_0=-\fft32r$&$r\le-\fft23$&anti-chiral\\
&&$e_0=-\fft32r+2$&$r<0$&SLI\\
&&$e_0=\fft32r+2$&$r>0$&SLII\\
\hline
Vector II &$\mathcal D(e_0+6,0,0;r)$ & --- & --- & ---\\
\hline
Vector III &$\mathcal D(e_0+3,0,0;r-2)$& $e_0=-\fft32r$&$r\le0$& anti-chiral \\
&& $e_0=-\fft32r+2$&$r\le\fft23$& SLI \\
\hline
Vector IV &$\mathcal D(e_0+3,0,0;r+2)$& $e_0=\fft32r$&$r\ge0$ & chiral \\
&& $e_0=\fft32r+2$&$r\ge-\fft23$& SLII \\
\hline
\end{tabular}
\caption{The generic shortening structure.  For a given $e_0$ and
$r$ satisfying the shortening condition, there may be an additional
degeneracy associated with the global symmetries of SE$_5$. The
conserved gravitinos are present only if the compactification
preserves $\ge$16 real supercharges. Vector Multiplet II is never
shortened. \label{tbl:KKshort}}
\end{table}

\section{The shortened spectrum of IIB supergravity on $Y^{p,q}$}
\label{sec:YpqSpect}

The scalar Laplacian on $Y^{p,q}$ and aspects of the spectrum have been investigated in
\cite{Berenstein:2005xa,Kihara:2005nt,Oota:2005mr,Enciso:2010ry,Chen:2012wr}.
Much of the difficulty in obtaining the full spectrum is due to the fact that, although
the Laplacian is separable, one ends up with a second order equation of Heun
type.  This arises because of the cubic function $q(y)=b-3y^2+2y^3$ that shows up
in the metric.  It turns out, however, that the Heun equation admits simple solutions
once the shortening conditions (\ref{eq:shortc}) are imposed.

We follow the general analysis of \cite{Kihara:2005nt} and refer the reader to that
reference for additional notation and conventions.  The isometry group of $Y^{p,q}$
is $\mathrm{SU}(2)_j\times \mathrm U(1)_\alpha\times\mathrm U(1)_r$, and there
are three commuting Killing vectors, which may be taken to be $\partial/\partial\phi$,
$\partial/\partial\psi$ and $\partial/\partial\alpha$.  This suggests that we look for solutions to
the scalar equation (\ref{eq:laplacian}) of the separable form
\begin{equation}
Y(y,\theta,\phi,\psi,\alpha)=e^{i(N_\phi\phi+N_\psi\psi+N_\alpha\alpha/l)}R(y)\Theta(\theta).
\end{equation}
Note that the $R$-charge is given by
\begin{equation}
r=2N_\psi-\fft{N_\alpha}{3l},\label{eq:r-chargeRelation}
\end{equation}
where
\begin{equation}\label{eq:lpq}
\fft1l=\fft{3q^2-2p^2+p\sqrt{4p^2-3q^2}}q.
\end{equation}
Here $\Theta$ satisfies the equation
\begin{equation}
\left[\fft1{\sin\theta}\fft\partial{\partial\theta}\sin\theta\fft\partial{\partial\theta}
-\fft1{\sin^2\theta}(N_\phi+N_\psi\cos\theta)^2+j(j+1)-N_\psi^2\right]\Theta=0,
\end{equation}
and the equation for $R$ can be transformed to the standard form of Heun's
equation.

We first consider the $\Theta$ equation.  This may be solved in terms of the Jacobi
polynomials $P_n^{(\alpha,\beta)}$
\begin{equation}
\Theta=\left(\sin\fft\theta2\right)^{|N_\phi+N_\psi|}\left(\cos\fft\theta2\right)^{|N_\phi-N_\psi|}
P_{j-|N_\phi+N_\psi|/2-|N_\phi-N_\psi|/2}^{(|N_\phi+N_\psi|,|N_\phi-N_\psi|)}(\cos\theta),
\end{equation}
where
\begin{equation}
j\ge\max(|N_\phi|,|N_\psi|),
\end{equation}
and either
\begin{equation}
\{j,N_\phi,N_\psi\}\in\mathbb Z\qquad\mbox{or}\qquad
\{j,N_\phi,N_\psi\}\in\mathbb Z+\ft12.
\label{eq:SU2reg}
\end{equation}
These conditions ensure regularity at $\theta=0$ and $\pi$.

For the $R(y)$ equation, it may be converted to a standard Heun form by taking
\cite{Kihara:2005nt}
\begin{equation}
x=\fft{y-y_1}{y_2-y_1},
\end{equation}
where $y_1$, $y_2$ and $y_3$ are the roots of the cubic $q(y)$ ordered from smallest to
largest.  (The physical range of $y$ is $y_1\le y\le y_2$.)  Then
\begin{equation}
R=x^{\alpha_1}(1-x)^{\alpha_2}(a-x)^{\alpha_3}h(x),
\end{equation}
where $h(x)$ satisfies Heun's equation
\begin{equation}
h''(x)+\left(\fft\gamma{x}+\fft\delta{x-1}+\fft\epsilon{x-a}\right)h'(x)+\fft{\alpha\beta x-k}{x(x-1)(x-a)}
h(x)=0.
\end{equation}
The exponents $\alpha_i$ are given by
\begin{eqnarray}
\alpha_1&=&\fft14\left|r-N_\alpha\left(p+q-\fft1{3l}\right)\right|,\nn\\
\alpha_2&=&\fft14\left|r-N_\alpha\left(-p+q-\fft1{3l}\right)\right|,\nn\\
\alpha_3&=&\fft14\left|r-N_\alpha\left(-2q+\fft2{3l}\right)\right|.
\label{eq:alphaidef}
\end{eqnarray}
The parameters in Heun's equation are
\begin{eqnarray}
&&\alpha=-\fft12e_0+\alpha_1+\alpha_2+\alpha_3,\qquad
\beta=2+\fft12e_0+\alpha_1+\alpha_2+\alpha_3,\nn\\
&&\gamma=1+2\alpha_1,\qquad\delta=1+2\alpha_2,\qquad\epsilon=1+2\alpha_3,
\end{eqnarray}
along with
\begin{eqnarray}
a&=&\fft{q+\sqrt{4p^2-3q^2}}{2q},\nn\\
k&=&(\alpha_1+\alpha_3)(1+\alpha_1+\alpha_3)-\alpha_2^2
+a\left[(\alpha_1+\alpha_2)(1+\alpha_1+\alpha_2)-\alpha_3^2\right]\nn\\
&&-\fft{p}q\left[\fft1{12}\left(1+\fft{q}p(1+a)\right)e_0(e_0+4)-j(j+1)
+\fft1{16}\left(\fft23\fft{N_\alpha}l-r\right)^2\right].
\end{eqnarray}

Before we proceed to examine the shortening conditions, note that constant solutions
$h(x)$ require $\alpha\beta=0$ and $k=0$.  For $e_0\ge0$, this reduces to
\begin{equation}
h(x)=1\qquad\Leftrightarrow\qquad\alpha=0,\quad k=0.
\end{equation}
We also note that a linear solution requires $(\alpha+1)(\beta+1)=0$, along with a more
complicated condition on $k$ that may be written as
\begin{equation}
(k+\gamma+\epsilon)(k+a(\gamma+\delta))-a\delta\epsilon=0.
\label{eq:linkcond}
\end{equation}

\subsection{The shortening condition $e_0=\fft32|r|$}

We first consider the case $e_0=\fft32r$.  Examining $\alpha$, we find
\begin{eqnarray}
2\alpha&=&-\fft32r+2(\alpha_1+\alpha_2+\alpha_3)\nn\\
&=&-\left(3N_\psi-N_\alpha\fft1{2l}\right)+\left|N_\psi-N_\alpha\fft{p+q}2\right|
+\left|N_\psi-N_\alpha\fft{-p+q}2\right|+\left|N_\psi-N_\alpha\left(-q+\fft1{2l}\right)\right|.\nn\\
\label{eq:2alphacond}
\end{eqnarray}
If we were to ignore the absolute values, then this expression simply yields $\alpha=0$.  This
suggests that $e_0=\fft32r$ shortening corresponds to constant solutions to Heun's equation.
Following through on this conjecture, we see that the absolute values are such that
$\alpha$ vanishes whenever
\begin{equation}
N_\psi\ge\max\left(N_\alpha\fft{p+q}2,-N_\alpha\fft{p-q}2,N_\alpha\left(\fft1{2l}-q\right)\right).
\end{equation}
Since the third quantity lies between the first two, it does not provide any further restriction
in the inequality.  Which of the first two quantities is greater depends on the sign of $N_\alpha$,
and we find
\begin{eqnarray}
&&N_\psi\ge N_\alpha\fft{p+q}2\kern2.6em\mbox{for}\quad N_\alpha\ge0;\nn\\
&&N_\psi\ge(-N_\alpha)\fft{p-q}2\quad\mbox{for}\quad N_\alpha\le0,
\label{eq:e0cond1}
\end{eqnarray}
as a necessary condition for obtaining a constant solution.

For a constant solution to exist, we must also demand $k=0$.  Assuming that $N_\psi$
satisfies the condition (\ref{eq:e0cond1}), which corresponds to simply dropping the
absolute values in (\ref{eq:alphaidef}), we find
\begin{equation}
k=\fft{p}q(j-N_\psi)(j+1+N_\psi).
\label{eq:heunk}
\end{equation}
In this case, $k=0$ corresponds to either $j=N_\psi$ or $j=-(N_\psi+1)$.  Since both
$j$ and $N_\psi$ are non-negative, we conclude that $j=N_\psi$ is required.  Putting
everything together then gives
\begin{equation}
\mbox{For } e_0=\ft32r:\qquad j=N_\psi\ge0\qquad\mbox{and}\qquad
-N_\psi\fft2{p-q}\le N_\alpha\le N_\psi\fft2{p+q}. \label{eq:e=3r/2}
\end{equation}
Note that, for a given $j$, we must also include the $(2j+1)$-fold
degeneracy of the SU(2) harmonics, corresponding to $-j\le N_\phi\le
j$.

We now consider how the U(1) quantum numbers $\{N_\phi,N_\psi,N_\alpha\}$ are quantized.
Based on the periodicities of the U(1) circles, all three quantities are integer spaced.  However,
there may be a shift imposed by regularity at the poles of the spheres in the $Y^{p,q}$ metric.
For $\theta=0$ and $\pi$, the regularity condition for the SU(2) harmonics was given in
(\ref{eq:SU2reg}).  However, we must also consider regularity at $y=y_1$ and $y_2$.  To
examine this, we note that the $\alpha$ circle is given by
\begin{equation}
d\left(\fft\alpha{l}\right)+\fft{b-2y+y^2}{6(b-y^2)}(d\psi-\cos\theta d\phi).
\end{equation}
At $y=y_1$, this becomes
\begin{equation}
d\left(\fft\alpha{l}\right)+\fft{p+q}2(d\psi-\cos\theta d\phi),
\end{equation}
while at $y=y_2$, this becomes
\begin{equation}
d\left(\fft\alpha{l}\right)-\fft{p-q}2(d\psi-\cos\theta d\phi).
\end{equation}
Thus at $y=y_1$, the natural U(1) coordinate is $\alpha/l+(p+q)\psi/2$, and at $y=y_2$,
the natural coordinate is $\alpha/l-(p-q)\psi/2$.  While $N_\alpha$ remains integral, this
shifts the quantization of $N_\psi$, depending on whether $p+q$ is even or odd.  For
$p+q$ even, we have
\begin{equation}
N_\alpha\in\mathbb Z\quad\mbox{and}\quad N_\psi\in\mathbb Z,
\label{eq:evencond}
\end{equation}
and for $p+q$ odd, we have
\begin{eqnarray}
&&N_\alpha\in2\mathbb Z\quad\mbox{and}\quad N_\psi\in\mathbb Z,\nn\\
\mbox{or}\qquad
&&N_\alpha\in2\mathbb Z+1\quad\mbox{and}\quad N_\psi\in\mathbb Z+\ft12.
\label{eq:oddcond}
\end{eqnarray}
Note that these quantization conditions apply not only to the
solutions with the shortening condition $e_0=\fft32|r|$, but also to
the ones with $e_0=\fft32|r|+2$ discussed below.

As noted in \cite{Berenstein:2005xa}, the constant solutions to Heun's equation correspond
to holomorphic functions on the Calabi-Yau cone.  Since these are the only functions that
saturate the bound $e_0\ge\fft32r$ \cite{Eager:2012hx}, we conclude that the identification
of constant solutions with the shortening condition $e_0=\fft32r$ is complete.

In the above, we have considered positive $R$-charge.   The negative $r$ modes may be
obtained by taking the complex conjugate.  The result is
\begin{equation}
\mbox{For } e_0=-\ft32r:\qquad j=-N_\psi\ge0\qquad\mbox{and}\qquad
N_\psi\fft2{p+q}\le N_\alpha\le-N_\psi\fft2{p-q}.
\end{equation}

Before proceeding to the other type of shortening, it is worth
noting that from (\ref{eq:e=3r/2}), (\ref{eq:evencond}) and
(\ref{eq:oddcond}), it is not difficult to convince oneself that,
except for $Y^{1,0}=T^{1,1}$, there are no shortened multiplets in
the spectrum of $Y^{p,q}$ manifolds that have $e_{0}<2$. Thus,
$T^{1,1}$ is the only member of the family of $Y^{p,q}$ manifolds
that has shortened multiplets with scalar fields that need to be
quantized with Neumann boundary conditions in AdS$_{5}$.

\subsection{The shortening condition $e_0=\fft32|r|+2$}

We now consider the case $e_0=\fft32r+2$.  Again, we start with $\alpha$.  Since $e_0$ is
increased by 2, the expression (\ref{eq:2alphacond}) becomes
\begin{equation}
2\alpha=-\left(2+3N_\psi-N_\alpha\fft1{2l}\right)+\left|N_\psi-N_\alpha\fft{p+q}2\right|
+\left|N_\psi-N_\alpha\fft{-p+q}2\right|+\left|N_\psi-N_\alpha\left(-q+\fft1{2l}\right)\right|.
\label{eq:2alphacond2}
\end{equation}
If we impose the conditions (\ref{eq:e0cond1}), we would now obtain $\alpha=-1$,
which suggests a linear solution to Heun's equation.  However, there is a second
possibility that $\alpha=0$, which gives a constant solution under the right conditions.

We first consider $\alpha=-1$.  Based on (\ref{eq:e0cond1}), the Heun parameters are
\begin{eqnarray}
&&\alpha=-1,\qquad\beta=3+3N_\psi-\fft{N_\alpha}{2l},\nn\\
&&\gamma=1+N_\psi-N_\alpha\fft{p+q}2,\qquad
\delta=1+N_\psi+N_\alpha\fft{p-q}2,\qquad
\epsilon=1+N_\psi-N_\alpha\left(-q+\fft1{2l}\right),\nn\\
\end{eqnarray}
Substituting this into (\ref{eq:linkcond}) then gives
\begin{equation}
\fft{p^2}{q^2}(j-N_\psi)(j-N_\psi-1)(j+1+N_\psi)(j+2+N_\psi)=0.
\end{equation}
Since $N_\psi\ge0$, this is solved by either $j=N_\psi$ or $j=N_\psi+1$.

The second possibility, $\alpha=0$, may be obtained by relaxing the conditions
(\ref{eq:e0cond1}), so that one of the arguments inside the absolute values in
(\ref{eq:2alphacond2}) becomes negative.  There are two possibilities, depending
on the sign of $N_\alpha$
\begin{eqnarray}
&&N_\alpha>0,\qquad N_\psi=N_\alpha\fft{p+q}2-1,\nn\\
\mbox{or}\qquad&&N_\alpha<0,\qquad N_\psi=-N_\alpha\fft{p-q}2-1.
\label{eq:special2}
\end{eqnarray}
Note that these cases must be restricted to $N_\psi\ge0$.  In particular, this indicates
that $|N_\alpha|\ge2$ for $Y^{1,0}$.  Furthermore, the negative $N_\alpha$ case must
be restricted to $N_\alpha\le-2$ for $Y^{p,p-1}$, and does not exist for $Y^{p,p}$.
In all cases, we find that $J=N_\psi$ is required to give $k=0$, and hence a
constant solution to Heun's equation.

Combining the $\alpha=0$ and $\alpha=-1$ possibilities, we find that the $e_0=\fft32r+2$
states are given by
\begin{eqnarray}
\mbox{For }e_0=\ft32r+2:\qquad&&i)\quad
j=N_\psi\ge0\qquad\!\mbox{and}\!\!\qquad
-(N_\psi+1)\fft2{p-q}\le N_\alpha\le (N_\psi+1)\fft2{p+q},\nn\\
&&ii)\quad
j=N_\psi+1\ge1\qquad\mbox{and}\qquad
-N_\psi\fft2{p-q}\le N_\alpha\le N_\psi\fft2{p+q}.\nn\\
\end{eqnarray}
The $e_0=-\fft32r+2$ states can be obtained by complex conjugation.

We have found that the shortening conditions $e_0=\fft32|r|$ and $e_0=\fft32|r|+2$
are satisfied by a combination of constant and linear solutions to Heun's equation.
For $e_0=\fft32|r|$, the solutions are complete, while for $e_0=\fft32|r|+2$
we have been unable to prove that these are the only possible solutions.  Nevertheless,
the evidence that we present below for the superconformal index and for $c-a$
strongly suggests that these solutions are complete.  We have also checked this
numerically for the low-lying spectrum with $j\le5$ and $e_0\le15$ for $Y^{p,q}$
with $0<q<p<10$.

These shortening conditions are summarized in Table~\ref{tbl:scalarshort}.  The
generic shortened spectrum of IIB supergravity on $Y^{p,q}$ (with $p>q>0$) is
then obtained by combining these scalar eigenvalues with the shortening structure
given in Table~\ref{tbl:KKshort}.  The complete shortened KK spectrum is given
by these generic towers along with Betti multiplets that are in one to one correspondence
to the ones
present in the case of $T^{1,1}$ \cite{Ceresole:1999zs,Ceresole:1999ht}.  This is
because $Y^{p,q}$ and $T^{1,1}$ share the same topology $S^2\times S^3$.  In
principle, the spectrum on a generic Sasaki-Einstein manifold may include special
multiplets.  However, these do not arise in the case of $Y^{p,q}$.  Finally, note that,
other than the Betti vector multiplet, there are four conserved
vector multiplets (one with $j=0$ and three with $j=1$) in the spectrum related to the
global isometry group $\mathrm{SU}(2)\times\mathrm U(1)$.

\begin{table}
\centering
\begin{tabular}{|l|l|l|l|}
\hline
Shortening&Condition for $j$&Condition for $N_\alpha$&Condition\\
&&&~~for $N_\psi$\\
\hline
$e_0=\fft32r$&$j=N_\psi$&$-N_\psi\fft2{p-q}\le N_\alpha\le N_\psi\fft2{p+q}$&$N_\psi\ge0$\\
\hline
$e_0=-\fft32r$&$j=-N_\psi$&$N_\psi\fft2{p+q}\le N_\alpha\le-N_\psi\fft2{p-q}$&$N_\psi\le0$\\
\hline
$e_0=\fft32r+2$&$j=N_\psi$&$-(N_\psi+1)\fft2{p-q}\le N_\alpha\le (N_\psi+1)\fft2{p+q}$&$N_\psi\ge0$\\
&$j=N_\psi+1$&$-N_\psi\fft2{p-q}\le N_\alpha\le N_\psi\fft2{p+q}$&$N_\psi\ge0$\\
\hline
$e_0=-\fft32r+2$&$j=-N_\psi$&$-(-N_\psi+1)\fft2{p+q}\le N_\alpha\le(-N_\psi+1)\fft2{p-q}$&$N_\psi\le0$\\
&$j=-N_\psi+1$&$N_\psi\fft2{p+q}\le N_\alpha\le-N_\psi\fft2{p-q}$&$N_\psi\le0$\\
\hline
\end{tabular}
\caption{The scalar eigenstates leading to the shortened spectrum on
$Y^{p,q}$.  The quantum numbers $N_\psi$ and $N_\alpha$ are
restricted to satisfy the conditions (\ref{eq:evencond}) and
(\ref{eq:oddcond}). $r$ is given by (\ref{eq:r-chargeRelation}).
Each state is $(2j+1)$-fold degenerate, corresponding to the allowed
values of $|N_\phi|\le j$. \label{tbl:scalarshort}}
\end{table}

\section{AdS/CFT state-operator correspondence in the mesonic chiral ring}
\label{sec:Ring}

The full mesonic chiral ring of the dual field theories were
constructed in \cite{Benvenuti:2005}. The gauge theory dual to
$Y^{p,q}$ was realized to have three types of mesonic blocks denoted
by $\mathcal{S}$, $\mathcal{L}_{+}$, and $\mathcal{L}_{-}$. The
product of $\mathcal{L}_{+}$ and $\mathcal{L}_{-}$ results in an
operator of the form $\mathcal{S}^{p}$. Thus a general mesonic
chiral BPS operator $\mathcal{O}_{s,N_{\alpha}}$ can be written as
\begin{equation}
\mathcal{O}_{s,N_{\alpha}}=\mathcal{S}^{s}\mathcal{L}^{N_{\alpha}},
\end{equation}
where $\mathcal{L}=\mathcal{L}_{+}$ and
$\mathcal{L}^{-1}=\mathcal{L}_{-}$. According to
\cite{Benvenuti:2005} the operator constructed in this way has an
R-charge
\begin{equation}
Q_{R}[\mathcal{O}_{s,N_{\alpha}}]=2s+p|N_{\alpha}|+N_{\alpha}\left(q-\frac{1}{3l}\right),\label{eq:operatorR-charge}
\end{equation}
and transforms in an irreducible SU(2) representation with spin $j$
\begin{equation}
j[\mathcal{O}_{s,N_{\alpha}}]=s+|N_{\alpha}|\frac{p}{2}+N_{\alpha}\frac{q}{2}.
\end{equation}

From (\ref{eq:r-chargeRelation}) and (\ref{eq:operatorR-charge}) it
is clear that one has to identify
\begin{equation}
N_{\psi}=s+|N_{\alpha}|\frac{p}{2}+N_{\alpha}\frac{q}{2},
\end{equation}
and therefore $N_{\psi}=j\ge0$. Then from positivity of $s$ the
range of $N_{\alpha}$ in (\ref{eq:e=3r/2}) follows. The quantization
conditions (\ref{eq:evencond}) and (\ref{eq:oddcond}) result by
demanding $s$ to be an integer.

This establishes that the full mesonic chiral ring of the quiver
gauge theory is dual to the supergravity KK states with
$e_{0}=\fft32r$ in the Vector I tower. Extending the analysis to all
the protected operators and shortened multiplets is straightforward
following \cite{Eager:2012hx}.

\section{Extension to $p=q$ and $q=0$}
\label{sec:SpecialCases}

The family of $Y^{p,q}$ manifolds, $p>q>0$, is often formally
extended to include the cases $p=q$ where $Y^{p,p}\equiv S^{5}/\mathbb{Z}_{2p}$
and $q=0$ where $Y^{p,0}\equiv T^{1,1}/\mathbb{Z}_{p}$. Although the $Y^{p,q}$ metric is
not well-defined for these values of $p$ and $q$, such assignments are
natural from the point of view of the toric diagrams
\cite{Martelli:2004}. Here we find that the shortened spectrum
presented above gives in fact the untwisted shortened spectrum on
$S^{5}/\mathbb{Z}_{2p}$ and $T^{1,1}/\mathbb{Z}_{p}$ as one
specializes to $p=q$ and $q=0$, respectively%
\footnote{That this must be the case is obvious if one combines the
AdS/CFT state-operator correspondence with the
observation that the protected operators of the dual field theories
are obtainable in the limiting cases $p=q$ and $q=0$.}.
The reason for indicating untwisted here is because $S^{5}/\mathbb{Z}_{2p}$ also has a
twisted sector that we will discuss in the next section.

The untwisted shortened spectrums on $S^{5}/\mathbb{Z}_{2p}$ and
$T^{1,1}/\mathbb{Z}_{p}$ were discussed in \cite{Ardehali:2013}. To
see that the shortened KK spectrum presented above is consistent
with \cite{Ardehali:2013} when $p=q$ or $q=0$, only a comparison of
the conventions is required.

For $p=q$, we get $l=1/2p$ and $r=2N_{\psi}-\fft{2p}{3}N_{\alpha}$.
Both $N_{\psi}$ and $N_{\alpha}$ have integer quantization in this
case since $p+q=2p$ is even. To make contact with the spectrum
presented in \cite{Ardehali:2013}, one has to identify $N_{\psi}$
and $p N_{\alpha}$ with $k/2$ and $-Q/2$, respectively, in
Table~3 of that work%
\footnote{Unlike in \cite{Ardehali:2013}, here we are using $Q$ for
the $q$-charge of the orbifold multiplets to avoid confusion with
the $q$ in $Y^{p,q}$. For a similar reason the KK level will be
denoted in the present paper by $L$, instead of $p$ which was used
in that work.}.

For $q=0$, we have $1/l=0$ and $r=2N_{\psi}$. To make contact with
the spectrum, as discussed in \cite{Ardehali:2013}, recall that
$T^{1,1}$ has a SU(2)$_{j}\times$SU(2)$_{l}\times$U(1)$_{r}$
isometry. The SU(2)$_{j}$ is broken to U(1)$_{j}$ by the orbifold
group $\mathbb{Z}_{p}$. After we identify $N_{\psi}(=j)$ with the
SU(2)$_{l}$ quantum number and $N_{\alpha}$ with the U(1)$_{j}$
quantum number, the spectrum given here matches the one discussed in
\cite{Ardehali:2013}. The fact that the quantization conditions of
$N_{\psi}$ and $N_{\alpha}$ are complicated according to
(\ref{eq:evencond}) or (\ref{eq:oddcond}), depending on $p$, is
related to the complicated expressions that were obtained for the
$\gamma_{j}^{(p)}$ in that work.

\section{The superconformal index for $Y^{p,q}$}
\label{sec:Index}

Knowledge of the shortened spectrum allows us to compute the
superconformal index of IIB supergravity on AdS$_{5}\times Y^{p,q}.$
Recall that the superconformal index for an $\mathcal N = 1$ SCFT is
defined as \cite{Kinney:2005ej,Romelsberger:2005eg}%
\footnote{As defined below, this is the ``right-handed" index. One
can also define a ``left-handed" index $\mathcal{I}^L$ in which one
replaces $r$ with $-r$ and swaps $s_1$ and $s_2$ in both
(\ref{eq:indexDef}) and the definition of $\delta.$ Also, we focus on the single-particle
index throughout this paper.}
\begin{equation}
\mathcal{I}^R=\mathrm{Tr}(-1)^{F}e^{-\beta\delta}t^{2(E+s_{2})}y^{2s_{1}},
\label{eq:indexDef}
\end{equation}
where $\delta=E-\fft32 r-2s_{2}$. Only states with $\delta = 0$ contribute to the index.
This condition means that only states which lie within shortened representations of the
superconformal algebra will contribute to the index. The index is therefore a protected
quantity and is independent of the coupling.
One may also refine the index to include chemical potentials for any global symmetries
of the theory by defining
\begin{equation}
\mathcal{I}^R=\mathrm{Tr}(-1)^{F}e^{-\beta\delta}t^{2(E+s_{2})}y^{2s_{1}}\prod a_i^{2q_i},
\label{eq:indexDef2}
\end{equation}
where the $a_i$ are exponentiated chemical potentials and the $q_i$ are charges under the
relevant symmetries.

For holographic theories, the Kaluza-Klein tower provides access to the spectrum of single trace
operators of the SCFT. We can thus use knowledge of the KK tower to compute the contribution
to the index from these single trace operators. The contribution to the index
from the different types of shortened supergravity multiplets is given in
Table~\ref{tbl:IndexContributions}.

\begin{table}
\centering
\begin{tabular}{|l|l|c|c|}
\hline
Shortening&Representation& $\mathcal{I}^R$&$\mathcal{I}^L$\\
\hline
conserved&$\mathcal D(E_0,s_1,s_2,r)$&$ (-1)^{2(s_1+s_2)+1} t^{3r+6s_2+6}\chi_{s_1}(y)$
&$(-1)^{2(s_1+s_2)+1} t^{-3r+6s_1+6}\chi_{s_2}(y)$\\
\hline
chiral&$\mathcal D(E_0,s_1,0,r)$&$ (-1)^{2s_1} t^{3r}\chi_{s_1}(y)$&$0$\\
\hline
anti-chiral&$\mathcal D(E_0,0,s_2,r)$&$0$&$(-1)^{2s_2} t^{-3r}\chi_{s_2}(y)$\\
\hline
SLI&$\mathcal D(E_0,s_1,s_2,r)$&$0$&$(-1)^{2(s_1+s_2)+1} t^{-3r+6s_1+6}\chi_{s_2}(y)$\\
\hline
SLII&$\mathcal D(E_0,s_1,s_2,r)$&$(-1)^{2(s_1+s_2)+1} t^{3r+6s_2+6}\chi_{s_1}(y)$&$0$\\
\hline
\end{tabular}
\caption{Contributions to the superconformal index from the various shortened multiplets, where $\chi_{j}(y)$ is the spin-$j$ SU(2) character as defined in (\ref{eq:SU2char}). While we focus on the right-handed index $\mathcal{I}^R$ in the text, we have included the contributions to the left-handed index $\mathcal{I}^L$ for completeness. \label{tbl:IndexContributions}}
\end{table}

In \cite{Gadde:2011}, the contribution to the index from single
trace operators was computed in the large $N$ limit of the quiver
gauge theories dual to AdS$_5\times Y^{p,q}$ using matrix model
techniques. Lacking knowledge of the shortened supergravity spectrum
for generic $Y^{p,q}$ a precise check with the supergravity result
was done only for the case of $T^{1,1},$ for which the results of
\cite{Gadde:2011} were shown to be in agreement with the
supergravity calculation of \cite{Nakayama:2006,Gadde:2011}.

Following this, \cite{Eager:2012hx} performed a general analysis for
quiver gauge theories dual to AdS$_5\times$ SE$_5$ for arbitrary
smooth Sasaki-Einstein manifolds. These authors understood the index
of a quiver gauge theory as the Euler characteristic of the cyclic
homology of Ginzburg's dg algebra associated to the quiver, and
related it to the Kohn-Rossi Cohomology of the dual internal
geometry. This established the gauge/gravity matching of the index
for arbitrary smooth Sasaki-Einstein manifolds. Specializing to the
case where the Calabi-Yau cone over the Sasaki-Einstein manifold is
toric (as is the case for the $Y^{p,q}$ geometries) they arrived at
relations that required only combinatorial computations to obtain
the index.\footnote{We would like to thank R. Eager for correspondence on this point.} 
However, an explicit expression for the index of the $Y^{p,q}$ theories computed
purely from supergravity was not presented. In the following we will
use our results for the shortened spectrum on AdS$_5\times Y^{p,q}$
to compute the supergravity result for these geometries and compare
to the explicit result of \cite{Gadde:2011}.

For the $Y^{p,q}$ theories we include chemical potentials $a_1$ and
$a_2$ corresponding to the $\mathrm{SU}(2)_j\times\mathrm U(1)_\alpha$ global
symmetries of $Y^{p,q}$.  The accompanying charges are given by
$q_1=j$ and $q_2=p N_\alpha/2.$ The chemical potential for the $\mathrm SU(2)_j$
has the effect of introducing a factor of $\chi_{j}(a_2),$ which is
the spin-$j$ SU(2) character given by
\begin{equation}\label{eq:SU2char}
\chi_{j}(x) = \sum^{2j}_{k=0} x^{2j-2k} = \frac{x^{2j+1}-x^{-(2j+1)}}{x-x^{-1}}.
\end{equation}

We now move to evaluate the index. Referring to
Table~\ref{tbl:IndexContributions}, we see that only conserved,
chiral and SLII multiplets contribute. As happens for the
holographic $c-a,$ the conserved multiplets contribute the same as
evaluating the contribution from a semi-long multiplet at the
appropriate level. Also note that the contribution from the Betti
multiplets vanishes as noted for $T^{1,1}$ in \cite{Nakayama:2006}.
To compute the index, we sum the contributions from all of the
relevant multiplets in Table~\ref{tbl:KKshort} over the allowed
values in Table~\ref{tbl:scalarshort}. Finally, we multiply each
contribution by
\begin{equation}
\frac{1}{(1-y^{-1}t^3)(1-y t^3)},\label{eq:IndexPrefactor}
\end{equation}
which takes into account the geometric sum arising from the
contribution of the infinite set of operators constructed by acting
with space-time derivatives on the bare operators.

After performing all of the sums we arrive, for $p>q$, at the
following result for the evaluated index
\begin{eqnarray}\label{eq:Ypqindex}
\mathcal{I}^R &=&  \frac{a_1^p a_2^{p+q} t^{3(q+p)-1/l}}{1-a_1^p a_2^{p+q} t^{3(q+p)-1/l}} + \frac{a_1^p a_2^{-p-q} t^{3(q+p)-1/l}}{1-a_1^p a_2^{-p-q} t^{3(q+p)-1/l}} \nn \\
&& + \frac{a_1^{-p} a_2^{p-q} t^{-3(q-p)+1/l}}{1-a_1^{-p} a_2^{p-q} t^{-3(q-p)+1/l}} + \frac{a_1^{-p} a_2^{-p+q} t^{-3(q-p)+1/l}}{1-a_1^{-p} a_2^{-p+q} t^{-3(q-p)+1/l}},
\end{eqnarray}
which, upon replacing $1/l$ with (\ref{eq:lpq}), precisely agrees with the result of \cite{Gadde:2011} for $\mathcal{I}^R.$ This provides evidence that the shortened spectrum we have obtained above is complete. Although this result was strictly derived for $p>q$, Eq.~(\ref{eq:Ypqindex}) also gives the correct expression for the index when $p=q>1$, which corresponds to $S^5/\mathbb Z_{2p}.$%
\footnote{The case $p=1$, corresponding to the $\mathbb Z_2$ orbifold, is a special case as it preserves $\mathcal N=2$ supersymmetry and the index receives additional contributions.}

The derivation of (\ref{eq:Ypqindex}) for the case of $p=q$ requires extra care for two
reasons. First, since these geometries are singular, there are
twisted sector modes that must be taken into account. Second,
unlike for other $Y^{p,q}$, here the
SLII multiplets of the Gravitino I, Vector I and Vector IV towers with
$N_{\alpha}<0$ have their $N_{\psi}$ bounded from below by zero,
instead of the generic expression $-\fft{p-q}{2}N_{\alpha}-1$ which would
give the unacceptable value of $-1$. A similar comment applies to the CP conjugate SLI
multiplets for which the generic higher bound of $+1$ would be unacceptable for
$N_{\psi}$.

These cases were partially investigated in \cite{Nakayama:2005},
where $Y^{1,1}\equiv S^{5}/\mathbb{Z}_{2}$ was completely dealt with.
For $Y^{p,p}$ with $p>1$, the contribution to the index from the
untwisted sector was compared with the field theory computation and
the difference was conjectured to come from the twisted modes. Here
we confirm the prediction made in \cite{Nakayama:2005} by an
explicit computation of the twisted sector contribution. However, in
the case of $S^{5}/\mathbb{Z}_{2}$, we find an answer that
disagrees with \cite{Nakayama:2005} but agrees with
\cite{Gadde:2009dj}%
\footnote{The index computed in \cite{Gadde:2009dj} is the $\mathcal N=2$ index appropriate for the $\mathbb Z_2$ orbifold theory which preserves $\mathcal N=2$ supersymmetry. Using our results, we find agreement with the twisted sector index of \cite{Gadde:2009dj} upon setting $v=1$ in that paper to reduce their result to the $\mathcal N=1$ index.}.
We refer the reader to
\cite{Gadde:2009dj} for additional discussion of the $Y^{1,1}$ case
and focus on $Y^{p,p}$ with $p>1$ in the following.

The twisted sector states of these theories were discussed in an
$\mathcal{N}=2$ language suitable for the SU(2,2$|$1) index
computation in \cite{Ardehali:2013}. These states arise from the KK
reduction on the $S^1$ of the $(2,0)$ theory on AdS$_{5}\times
S^{1}$. The twisted states of $Y^{1,1}$ are shown in
Table~\ref{tbl:twisted}. To obtain the twisted states of $Y^{p,p}$
one keeps only the states with $\mathrm{U}(1)_Q=0 \mathrm{\ mod\
}2p$.

\begin{table}[t]
\centering
\begin{tabular}{|l|l|l|}
\hline
KK level&Representation&Shortening type\\
\hline
$L=0$&$\mathcal D(2,0,0;0)\mathbf1_0$&conserved\\
&$\mathcal D(2,0,0;\ft43)\mathbf1_{-2}$&chiral\\
&$\mathcal D(2,0,0;-\ft43)\mathbf1_2$&anti-chiral\\
$L\ge1$&$\mathcal D(L+1,0,0;\ft23(L+1))\mathbf1_{2L+2}$&chiral\\
&$\mathcal D(L+\ft32,\ft12,0;\ft23(L+\ft32))\mathbf1_{2L}$&\\
&$\mathcal D(L+2,0,0;\ft23(L+2))\mathbf1_{2L-2}$&\\
&$\mathcal D(L+1,0,0;-\ft23(L+1))\mathbf1_{-2L-2}$&anti-chiral\\
&$\mathcal D(L+\ft32,0,\ft12;-\ft23(L+\ft32))\mathbf1_{-2L}$&\\
&$\mathcal D(L+2,0,0;-\ft23(L+2))\mathbf1_{-2L+2}$&\\
\hline
\end{tabular}
\caption{\label{tbl:twisted}The twisted sector states for the
orbifold $S^5/\mathbb Z_2$ written in an $\mathcal N=2$ language. We
use the same $\mathrm{SU}(2)\times\mathrm U(1)_Q \times\mathrm
U(1)_r$ decomposition as in \cite{Ardehali:2013}. The restriction to
states with $\mathrm{U}(1)_Q=0 \mathrm{\ mod\ }2p$ yields the
twisted states of $Y^{p,p}$.}
\end{table}

The states with U(1)$_{Q}=0$ can be regarded as Betti multiplets and
give canceling contributions to the index. The chiral multiplets
give contributions according to Table~\ref{tbl:IndexContributions}.
Adding everything up and multiplying by the prefactor
(\ref{eq:IndexPrefactor}) yields for $p>1$
\begin{equation}\label{eq:Ypptwist}
\mathcal{I}^{R}_{Y^{p,p}\mbox{\small{\
twisted}}}=\frac{t^{2p}}{1-t^{2p}}.
\end{equation}
This coincides with the conjecture made in \cite{Nakayama:2005} for
$S^{5}/\mathbb{Z}_{2p}$ with $\mathbb{Z}_{2p}\subset\mathrm{SU}(3)$
generated by
\begin{equation}
\Omega=\begin{pmatrix} \omega\cr &\omega\cr
&&\omega^{-2}\end{pmatrix}, \label{eq:Z2pAct}
\end{equation}
with $\omega^{2p}=1$. In particular, adding the result (\ref{eq:Ypptwist}) from the twisted sector to the result of \cite{Nakayama:2005}
for the untwisted sector one recovers (\ref{eq:Ypqindex}) evaluated at $p=q>1$ with $a_1=a_2=1$.

\section{Holographic $c-a$ for $Y^{p,q}$ and a $1/N^{2}$ test of AdS/CFT}
\label{sec:c-a}

The quiver CFTs dual to IIB supergravity on AdS$_{5}\times Y^{p,q}$
have $2p$ nodes and a number of chiral fields in bifundamentals
\cite{Benvenuti:2004}.  A simple computation demonstrates that each
node in the quiver contributes a factor of $1/16$ to $c-a$ that is related to
the presence of a decoupled U(1).  Our aim is to reproduce this result
$c-a=2p/16=p/8$ from the gravitational dual using the shortened KK spectrum.

As discussed in \cite{Ardehali:2013}, the holographic formula for
$c-a$ is \cite{Mansfield:2000zw,Mansfield:2002pa,Mansfield:2003gs}
\begin{equation}
c-a=-\fft1{360}\sum(-1)^F(E_0-2)d(s_1,s_2)\left(1+f(s_1)+f(s_2)\right),
\label{eq:c-aexpr}
\end{equation}
where the sum is in principle over all fields in the spectrum, but
can be restricted to only the shortened representations in the KK
tower, as the contribution from long multiplets automatically
vanish. Here, $E_{0}$, $s_1$ and $s_2$ are the quantum numbers of
the maximal compact subgroup of SO(4,2) labeling the bulk fields,
$d(s_1,s_2)=d(s_1)d(s_2)=(2s_1+1)(2s_2+1)$ is the dimension of the
$\mathrm{SO}(4)\simeq\mathrm{SU}(2)\times\mathrm{SU}(2)$
representation and $f(X)=X(X+1)(6X(X+1)-7)$.

The sum in (\ref{eq:c-aexpr}) is divergent and requires
regularization. Following
\cite{Mansfield:2002pa,Mansfield:2003gs,Ardehali:2013gra,Ardehali:2013},
we regularize the sum by multiplying each term by $z^{L}$, where $L$
is the `KK level' of the bulk field, and then keep the finite part
of the resulting function of $z$ as $z\to 1$. Since all the fields
in the same multiplet have the same KK level this regularization
manifestly preserves supersymmetry. As explained in
\cite{Ardehali:2013}, for a generic Sasaki-Einstein manifold there
is no well-defined notion of a KK level. Nevertheless one can assign
the bulk KK multiplets such levels $L$ according to what they would
have been had the multiplets come from compactification on $S^5$.
The level assignment is as follows.  We choose
\begin{equation}
L=E_{0}-s_{1}-s_{2},
\end{equation}
for the Graviton, Gravitino I, Gravitino III and Vector I towers,
and
\begin{equation}
L=E_{0}-s_{1}-s_{2}-1,
\end{equation}
for the Gravitino II, Gravitino IV, Vector III and Vector IV towers.
$E_0$ is again the lowest AdS energy eigenvalue (corresponding to
the conformal dimension $\Delta$ in the CFT dual) in the multiplet.
These level assignments for the shortened multiplets along with their
contribution to the holographic $c-a$ are listed in
Table~\ref{tbl:LevelsAndC-A}.

\begin{table}[tp]
\centering
\begin{tabular}{|l|l|l|l|l|}
\hline
Multiplet&Representation&Shortening&Level&$c-a$ for one\\
\hline
Graviton&$\mathcal D(e_0+3,\fft12,\fft12;r)$&conserved&$L=e_{0}+2$&$-\fft58$\\
&&SLI&&$-\fft5{48}(e_{0}+3)$\\
&&SLII&&$-\fft5{48}(e_{0}+3)$\\
\hline
Gravitino I&$\mathcal D(e_0+\fft32,\fft12,0;r+1)$&(conserved)&$L=e_{0}+1$&$(\fft{35}{192})$\\
&&chiral&&$-\fft5{48}e_{0}$\\
&&SLI&&$-\fft1{96}(e_{0}+\fft32)$\\
&&SLII&&$\fft5{48}(e_{0}+1)$\\
\hline
Gravitino II&$\mathcal D(e_0+\fft92,\fft12,0;r-1)$&SLI&$L=e_{0}+3$&$-\fft1{96}(e_{0}+\fft92)$\\
\hline
Gravitino III&$\mathcal D(e_0+\fft32,0,\fft12;r-1)$&(conserved)&$L=e_{0}+1$&$(\fft{35}{192})$\\
&&anti-chiral&&$-\fft5{48}e_{0}$\\
&&SLII&&$-\fft1{96}(e_{0}+\fft32)$\\
&&SLI&&$\fft5{48}(e_{0}+1)$\\
\hline
Gravitino IV &$\mathcal D(e_0+\fft92,0,\fft12;r+1)$&SLII&$L=e_{0}+3$&$-\fft1{96}(e_{0}+\fft92)$\\
\hline
Vector I &$\mathcal D(e_0,0,0;r)$&conserved&$L=e_{0}$&$\fft1{32}$\\
&&chiral&&$-\fft1{96}(e_{0}-\fft32)$\\
&&anti-chiral&&$-\fft1{96}(e_{0}-\fft32)$\\
&&SLI&&$\fft1{96}(e_{0}-\fft12)$\\
&&SLII&&$\fft1{96}(e_{0}-\fft12)$\\
\hline
Vector II &$\mathcal D(e_0+6,0,0;r)$ & --- & --- &\\
\hline
Vector III &$\mathcal D(e_0+3,0,0;r-2)$& anti-chiral &$L=e_{0}+2$&$-\fft1{96}(e_{0}+\fft32)$\\
&&SLI&&$\fft1{96}(e_{0}+\fft52)$ \\
\hline
Vector IV &$\mathcal D(e_0+3,0,0;r+2)$& chiral &$L=e_{0}+2$&$-\fft1{96}(e_{0}+\fft32)$\\
&&SLII&&$\fft1{96}(e_{0}+\fft52)$ \\
\hline
\end{tabular}
\caption{Level assignment and the contribution of a single shortened
multiplet to $c-a$ for the shortened spectrum of
Table~\ref{tbl:KKshort}. The latter must be multiplied by the
degeneracy of the multiplet, which depends on the compactification
manifold. The conserved gravitinos are present only if the
compactification preserves $\ge$16 real supercharges, hence absent
from all $Y^{p,q}$ except $Y^{1,1}$. \label{tbl:LevelsAndC-A}}
\end{table}

The above regularization has proven successful for
$S^{5}/\mathbb{Z}_{n}$ and $T^{1,1}/\mathbb{Z}_{n}$ with $n$ an
arbitrary natural number \cite{Ardehali:2013}. Here we find that it
continues to be successful for all $Y^{p,q}$ manifolds with $p>q>0$.
The details of the computation are quite similar to the cases
discussed in \cite{Ardehali:2013}. As an example, we present the
computation of $c-a$ for the graviton tower. We have
\begin{eqnarray}
c-a\Big|_{\rm
graviton}&=&-\fft58+2\times\!\!\sum_{N_{\alpha},N_{\psi},N_{\phi}} z^{e_{0}+2}
\left(-\fft{5}{48}\right)(e_{0}+3)\nn \\
&=&-\fft58+2\times\!\!\sum_{N_{\alpha},N_{\psi}}
z^{\fft32(2N_{\psi}-{N_{\alpha}}/{3l})+2}\left(-\frac{5}{48}\right)
\left(\fft32\left(2N_{\psi}-\fft{N_{\alpha}}{3l}\right)+3\right)(2N_{\psi}+1).\nn\\
\label{eq:c-aGraviton}
\end{eqnarray}
We first sum over $N_{\psi}$ and then over $N_{\alpha}$. This is facilitated
by taking the three cases $N_{\alpha}=0$, $N_{\alpha}>0$ and
$N_{\alpha}<0$ separately. For $N_{\alpha}=0$, the sum over
$N_{\psi}$ should go from $1$ to infinity, whereas for
$N_{\alpha}>0$ the sum over $N_{\psi}$ should go from
$\frac{p+q}{2}N_{\alpha}$ to infinity, and finally for
$N_{\alpha}<0$ the sum over $N_{\psi}$ should go from
$-\frac{p-q}{2}N_{\alpha}$ to infinity. Note that (as in the
computation of the index) while $N_{\alpha}$ is always an integer,
$N_{\psi}$ can be integer or half integer depending on
$N_{\alpha}/(p+q)$, but it always increases in unit steps for any
fixed $N_{\alpha}$. Evaluating the sum and expanding around $z=1$
gives
\begin{equation}
\begin{split}
c-a\Big|_{\rm
graviton}=&\fft{\alpha_{4}}{(z-1)^{4}}+\fft{\alpha_{3}}{(z-1)^{3}}+
\fft{\alpha_{2}}{(z-1)^{2}}+\alpha_{0}+\cdots,
\end{split}
\end{equation}
where $\alpha_{0,2,3,4}$ are complicated functions of $p$ and $q$ and
the ellipsis denotes terms vanishing as $z\to1$.

The contributions from the other towers can be worked out in a
similar manner. In addition, the contribution from the Betti vector
($1/32$) cancels against that from the Betti hyper ($-1/32$).
Putting everything together, one arrives at
\begin{equation}
c-a\Big|_{Y^{p,q}}=\fft{\alpha(Y^{p,q})}{(z-1)^2}+\fft{\alpha(Y^{p,q})}{(z-1)}+\fft{p}{8}+\cdots,
\label{eq:c-aYpq}
\end{equation}
where
\begin{equation}
\alpha(Y^{p,q})=\frac{q^{2}\left(4p-\sqrt{4p^{2}-3q^{2}}\right)}{4p^{2}\left(2p^{2}-3q^{2}-p\sqrt{4p^{2}-3q^{2}}\right)}.
\label{eq:c-aYpqPole}
\end{equation}
This computation is valid for $p>q$.

While the fourth and third order poles cancel at $z=1$, the second and first order poles do not.
This is similar to what happened for $Y^{p,p}$ and $Y^{p,0}$ that were considered in
\cite{Ardehali:2013}.  Following
\cite{Mansfield:2002pa,Ardehali:2013gra,Ardehali:2013}, we drop the
pole terms, so we are left with the finite result $c-a=p/8$ for
$Y^{p,q}$, in perfect agreement with the field theory result.

\section{Discussion}

We conjecture that the shortened spectrum of the scalar Laplacian
presented in section~\ref{sec:YpqSpect} is complete.  While we have no proof
that Table~\ref{tbl:scalarshort} gives the complete shortened spectrum of IIB
supergravity on $Y^{p,q}$, the successful computation of the superconformal index
in section~\ref{sec:Index} gives us confidence that the shortened spectrum is in fact
fully obtained.

Using the shortened spectrum, we holographically computed $c-a$
of the corresponding dual field theories and found agreement. The computation
involves a regularization that we advocated in \cite{Ardehali:2013}.
It is not entirely clear to us why this is the correct procedure
to use, but it had been successful for orbifolds of $S^{5}$
and $T^{1,1}$. The fact that this regularization succeeds for
$Y^{p,q}$ makes us believe that the method is correct and the
matchings found previously
\cite{Mansfield:2003gs,Ardehali:2013gra,Ardehali:2013} were not
coincidental. Thus we are able to restate some of the conclusions
made in \cite{Ardehali:2013} more confidently:
\begin{itemize}
\item  At least for the models we have considered (four-dimensional CFTs dual to
IIB string theory on AdS$_{5}\times X^5$, with $X^5$
either an orbifold of $S^5$, an orbifold of $T^{1,1}$ or
$Y^{p,q}$), holographic $c-a$ can be obtained purely from ten
dimensional supergravity, and as such, does not necessitate a stringy
origin. Massive string loop (or $\alpha'$) corrections to
holographic $c-a$ must vanish, as opposed to claims made to the
contrary in \cite{Liu:2010gz} and \cite{Anselmi:1998zb}.
\item The shift in holographic $c-a$ due to alternative boundary conditions (or the shift
in $c-a$ due to a relevant double-trace deformation) must vanish (at
least in the case of $T^{1,1}$), as opposed to the claim
made to the contrary in \cite{Nolland:2003kc}.
\end{itemize}

Another observation one can make is that holographic $c-a$ as a
function of $z$ can be thought of as an index, since it is computed
only from the shortened spectrum. It is of course most interesting
when it is expanded around $z=1$, where the constant term gives the
actual field theoretical $c-a$. But one can also pay attention to
the pole terms near $z=1$. Eq.~(\ref{eq:c-aYpq}) shows that
there are first and second order poles in the expansion of this
`index', with equal coefficients. This also happens for the even
orbifolds of $S^5$, where the pole coefficients can interestingly%
\footnote{Note that no twisted states have entered the derivation of
equation (\ref{eq:c-aYpq}). It seems that for $p>1$ the twisted
states of $S^{5}/\mathbb{Z}_{2p}$ precisely make up for the states
lost by setting $p=q$ in the spectrum of $Y^{p,q}$.}
be obtained from (\ref{eq:c-aYpqPole}) by setting $p=q$. However,
there are no pole terms for the odd orbifolds of $S^5$
\cite{Ardehali:2013}. Since the pole terms are related to the
asymptotic growth of the terms in the sum (\ref{eq:c-aexpr}), guided
by the analysis of \cite{Eager:2010dk} and the above special cases,
we conjecture that the general expression for the coefficient of the
pole terms is
\begin{equation}
\alpha(\mbox{SE}_{5})=-\frac{5}{32\pi^3}\left(\frac{\int_{\mbox{\scriptsize{SE}}_{5}}\mbox{Riem}^2}{40}-\mbox{vol}(\mbox{SE}_{5})\right).
\label{eq:c-aSE5Pole}
\end{equation}

More interesting than (\ref{eq:c-aSE5Pole}) would be a geometrical
expression for the constant piece in the expansion of the $c-a$
index, i.e.\ the field theoretical $c-a$. This remains a challenge,
perhaps, until a better understanding of the required regularization
method is gained. Possible relations between holographic $c-a$ and the
superconformal index are also currently under investigation.

Finally, note that
in field theory $c-a$ is proportional to $\text{tr} R$ which arises
in the U(1)$_R$ 't~Hooft anomaly from the
U(1)-gravitational-gravitational triangle diagram. The expression
(\ref{eq:c-aexpr}) for the holographic $c-a$
seems to provide an alternative explicit field theoretical formula
for this strictly in terms of
the quantum numbers of protected single trace operators. It would be
interesting to explore this relation further.

\acknowledgments

We would like to thank I.~Bah, R.~Eager, K.~Intriligator, J.~McGreevy and L.~Rastelli
for interesting conversations and correspondences.
This work was supported in part by the US Department of
Energy under grants DE-SC0007859 and DE-SC0007984.



\begin{thebibliography}{99}

\bibitem{Gauntlett:2004a}
J.~P.~Gauntlett, D.~Martelli, J.~Sparks and D.~Waldram, {\sl
Supersymmetric AdS$_5$ solutions of M-theory}, Class.\ Quant.\ Grav.
{\bf 21}, 4335 (2004). [hep-th/0402153]

\bibitem{Gauntlett:2004b}
J.~P.~Gauntlett, D.~Martelli, J.~Sparks and D.~Waldram, {\sl
Sasaki-Einstein metrics on $S^2 \times S^3$}, Adv.\ Theor.\ Math.\
Phys. {\bf 8}, 711 (2004). [hep-th/0403002]

\bibitem{Martelli:2004}
D.~Martelli and J.~Sparks, {\sl Toric Geometry, Sasaki-Einstein
Manifolds and a New Infinite Class of AdS/CFT Duals}, Commun.\
Math.\ Phys.\ {\bf 262} (2006) 51 [arXiv:0411238 [hep-th]].

\bibitem{Benvenuti:2004}
S.~Benvenuti, S.~Franco, A.~Hanany, D.~Martelli and J.~Sparks, {\sl
An Infinite Family of Superconformal Quiver Gauge Theories with
Sasaki-Einstein Duals}, JHEP {\bf 0506} (2005) 64
[arXiv:hep-th/0411264].

\bibitem{Canoura:2005}
F.~Canoura, J.~D.~Edelstein, L.~A.~P.~Zayas, A.~V.~Ramallo and
D.~Vaman, {\sl Supersymmetric branes on AdS$_5 \times Y^{p,q}$ and
their field theory duals}, JHEP {\bf 0603} (2006) 101
[arXiv:hep-th/0512087].

\bibitem{Benvenuti:2005}
S.~Benvenuti and M.~Kruczenski, {\sl Semiclassical strings in
Sasaki-Einstein manifolds and long operators in N=1 gauge theories},
JHEP {\bf 0610} (2006) 051 [arXiv:0505046 [hep-th]].

\bibitem{Gadde:2011}
A.~Gadde, L.~Rastelli, S.~S.~Razamat, and W.~Yan, {\sl On the
superconformal index of $\mathcal N = 1$ IR fixed points. A
holographic check}, JHEP {\bf 1103} (2011) 1 [arXiv:1011.5278
[hep-th]].

\bibitem{Berenstein:2005xa}
D.~Berenstein, C.~P.~Herzog, P.~Ouyang and S.~Pinansky,
{\sl Supersymmetry Breaking from a Calabi-Yau Singularity},
JHEP {\bf 0509} (2005) 084 [hep-th/0505029].

\bibitem{Kihara:2005nt}
H.~Kihara, M.~Sakaguchi and Y.~Yasui, {\sl Scalar Laplacian on
Sasaki-Einstein Manifolds $Y^{p,q}$}, Phys.\ Lett.\ B {\bf 621}
(2005) 288 [hep-th/0505259].

\bibitem{Oota:2005mr}
T.~Oota and Y.~Yasui, {\sl Toric Sasaki-Einstein Manifolds and Heun
Equations}, Nucl.\ Phys.\ B {\bf 742} (2006) 275 [hep-th/0512124].

\bibitem{Ardehali:2013}
A.~A.~Ardehali, J.~T.~Liu and P.~Szepietowski, {\sl $1/N^{2}$
corrections to the holographic Weyl anomaly},
JHEP {\bf 1401} (2014) 002 [arXiv:1310.2611 [hep-th]].

\bibitem{Gunaydin:1984fk}
M.~Gunaydin and N.~Marcus,
{\sl The spectrum of the $S^5$ compactification of the chiral $N=2$, $D=10$ supergravity
and the unitary supermultiplets of $U(2, 2/4)$},
Class.\ Quant.\ Grav.\  {\bf 2}, L11 (1985).

\bibitem{Kim:1985ez}
H.~J.~Kim, L.~J.~Romans and P.~van Nieuwenhuizen,
{\sl Mass spectrum of chiral ten-dimensional $N=2$ supergravity on $S^5$},
Phys.\ Rev.\ D {\bf 32}, 389 (1985).

\bibitem{Ceresole:1999zs}
A.~Ceresole, G.~Dall'Agata, R.~D'Auria and S.~Ferrara,
\emph{Spectrum of type IIB supergravity on AdS$_5\times T^{11}$:
Predictions on $\mathcal N=1$ SCFT's},
Phys.\ Rev.\ D {\bf 61}, 066001 (2000) [hep-th/9905226].

\bibitem{Ceresole:1999ht}
A.~Ceresole, G.~Dall'Agata and R.~D'Auria,
\emph{KK spectroscopy of type IIB supergravity on AdS$_5\times T^{11}$},
JHEP {\bf 9911}, 009 (1999) [hep-th/9907216].

\bibitem{Eager:2012hx}
R.~Eager, J.~Schmude and Y.~Tachikawa,
{\sl Superconformal Indices, Sasaki-Einstein Manifolds, and Cyclic Homologies},
arXiv:1207.0573 [hep-th].

\bibitem{Cassani:2010uw}
D.~Cassani, G.~Dall'Agata and A.~F.~Faedo, {\sl Type IIB
Supergravity on Squashed Sasaki-Einstein Manifolds}, JHEP {\bf 1005}
(2010) 094 [arXiv:1003.4283 [hep-th]].

\bibitem{Liu:2010sa}
J.~T.~Liu, P.~Szepietowski and Z.~Zhao, {\sl Consistent Massive
Truncations of IIB Supergravity on Sasaki-Einstein Manifolds},
Phys.\ Rev.\ D {\bf 81} (2010) 124028 [arXiv:1003.5374 [hep-th]].

\bibitem{Gauntlett:2010vu}
J.~P.~Gauntlett and O.~Varela, {\sl Universal Kaluza-Klein
Reductions of Type IIB to ${\mathcal{N}}\!=4$ Supergravity in Five
Dimensions}, JHEP {\bf 1006} (2010) 081 [arXiv:1003.5642 [hep-th]].

\bibitem{Skenderis:2010vz}
K.~Skenderis, M.~Taylor and D.~Tsimpis, {\sl A Consistent Truncation
of IIB Supergravity on Manifolds Admitting a Sasaki-Einstein
Structure}, JHEP {\bf 1006} (2010) 025 [arXiv:1003.5657 [hep-th]].

\bibitem{Flato:1983te}
M.~Flato and C.~Fronsdal,
\emph{Representations Of Conformal Supersymmetry},
Lett.\ Math.\ Phys.\  {\bf 8}, 159 (1984).

\bibitem{Dobrev:1985qv}
V.~K.~Dobrev and V.~B.~Petkova,
\emph{All Positive Energy Unitary Irreducible Representations of Extended Conformal
Supersymmetry},
Phys.\ Lett.\ B {\bf 162}, 127 (1985).

\bibitem{Freedman:1999gp}
D.~Z.~Freedman, S.~S.~Gubser, K.~Pilch and N.~P.~Warner,
\emph{Renormalization group flows from holography---supersymmetry and a c-theorem},
Adv.\ Theor.\ Math.\ Phys.\  {\bf 3}, 363 (1999) [hep-th/9904017].

\bibitem{Enciso:2010ry}
A.~Enciso and N.~Kamran,
{\sl Global Causal Propagator for the Klein-Gordon Equation on a Class of
Supersymmetric AdS Backgrounds},
Adv.\ Theor.\ Math.\ Phys.\ {\bf 14} (2010) 1183 [arXiv:1001.2200 [math-ph]].

\bibitem{Chen:2012wr}
F.~Chen, K.~Dasgupta, A.~Enciso, N.~Kamran and J.~Seo,
{\sl On the Scalar Spectrum of the $Y^{p,q}$ Manifolds},
JHEP {\bf 1205} (2012) 009 [arXiv:1201.5394 [hep-th]].

\bibitem{Romelsberger:2005eg}
C.~Romelsberger, {\sl Counting chiral primaries in N = 1, d=4
superconformal field theories}, Nucl.\ Phys.\ B {\bf 747}, 329
(2006) [hep-th/0510060].

\bibitem{Kinney:2005ej}
J.~Kinney, J.~M.~Maldacena, S.~Minwalla and S.~Raju, {\sl An Index
for 4 dimensional super conformal theories}, Commun.\ Math.\ Phys.\
{\bf 275}, 209 (2007) [hep-th/0510251].

\bibitem{Nakayama:2006}
Y.~Nakayama, {\sl Index for Supergravity on $AdS_5 \times T^{1,1}$
and Conifold Gauge Theory}, Nucl.\ Phys.\ B\ {\bf 755} (2006) 295
[arXiv:0602284 [hep-th]].

\bibitem{Nakayama:2005}
Y.~Nakayama, {\sl Index for Orbifold Quiver Gauge Theories}, Phys.\
Lett.\ B\ {\bf 636} (2006) 132 [arXiv:0512280 [hep-th]].

\bibitem{Gadde:2009dj}
A.~Gadde, E.~Pomoni and L.~Rastelli, {\sl The Veneziano Limit of
${\mathcal{N}}\!=2$ Superconformal QCD: Towards the String Dual of
${\mathcal{N}}\!=2$ $SU(N_c)$ SYM with $N_f = 2 N_c$},
arXiv:0912.4918 [hep-th].

\bibitem{Mansfield:2000zw}
P.~Mansfield and D.~Nolland, {\sl Order $1 / N^2$ Test of the
Maldacena Conjecture: Cancellation of the One Loop Weyl Anomaly},
Phys.\ Lett.\ B {\bf 495} (2000) 435 [hep-th/0005224].

\bibitem{Mansfield:2002pa}
P.~Mansfield, D.~Nolland and T.~Ueno, {\sl Order $1 / N^2$ Test of
the Maldacena Conjecture. 2. the Full Bulk One Loop Contribution to
the Boundary Weyl Anomaly}, Phys.\ Lett.\ B {\bf 565} (2003) 207
[hep-th/0208135].

\bibitem{Mansfield:2003gs}
P.~Mansfield, D.~Nolland and T.~Ueno, {\sl The Boundary Weyl Anomaly
in the ${\mathcal{N}}\!=4$ SYM / Type IIB Supergravity
Correspondence}, JHEP {\bf 0401} (2004) 013 [hep-th/0311021].

\bibitem{Ardehali:2013gra}
A.~A.~Ardehali, J.~T.~Liu and P.~Szepietowski, {\sl The Spectrum of
IIB supergravity on $\mathrm{AdS}_5\times S^5/\mathbb Z_3$ and a
$1/N^2$ test of AdS/CFT}, JHEP {\bf 1306} (2013) 024
[arXiv:1304.1540 [hep-th]].

\bibitem{Liu:2010gz}
J.~T.~Liu and R.~Minasian, {\sl Computing $1/N^2$ Corrections in
AdS/CFT}, arXiv:1010.6074 [hep-th].

\bibitem{Anselmi:1998zb}
D.~Anselmi and A.~Kehagias, {\sl Subleading Corrections and Central
Charges in the AdS / CFT Correspondence}, Phys.\ Lett.\ B {\bf 455}
(1999) 155 [hep-th/9812092].

\bibitem{Nolland:2003kc}
D.~Nolland, {\sl AdS / CFT Boundary Conditions, Multitrace
Perturbations, and the $c$-Theorem}, Phys.\ Lett.\ B {\bf 584}
(2004) 192 [hep-th/0310169].

\bibitem{Eager:2010dk}
R.~Eager, M.~Gary and M.~M.~Roberts,
{\sl Can you hear the shape of dual geometries?},
JHEP {\bf 1310}, 209 (2013) [arXiv:1011.5231 [hep-th]].

\end{thebibliography}
\end{document}